%% file: main.tex
\documentclass[%
    preprintnumbers,
    amsmath,
    amssymb,
    aps,
    physrev,
]{revtex4-2}

\usepackage{graphicx}
\usepackage{dcolumn}
\usepackage{bm}
\usepackage{hyperref}
\usepackage[capitalise]{cleveref}
\usepackage{caption}
\usepackage{subcaption}
\usepackage{float}
\usepackage{xstring}
\usepackage{xcolor}
\hypersetup{
    colorlinks=true,      
    linkcolor=blue,       
    citecolor=blue,      
    urlcolor=blue      
}
\newcommand{\be}{\begin{equation}}
\newcommand{\ee}{\end{equation}}
\newcommand{\ba}{\begin{eqnarray}}
\newcommand{\ea}{\end{eqnarray}}
\newcommand{\bd}{\begin{displaymath}}
\newcommand{\ed}{\end{displaymath}}

\crefname{section}{Sec.}{Secs.} 
\Crefname{section}{Section}{Sections} 

\newcommand{\Cite}[1]{%
\IfSubStr{#1}{,}{Refs.~\cite{#1}}{Ref.~\cite{#1}}%
}

\begin{document}
    \title{Variations of the crossover and first-order phase transition curve in modeling the QCD equation of state}
    \author{Joseph I. Kapusta}
    \email{kapusta@umn.edu}
    \author{Shensong Wan}
    \email{wan00589@umn.edu}
    \affiliation{School of Physics \& Astronomy, University of Minnesota, Minneapolis, 55455, MN, USA}

    \date{\today}


    \input{abstract}
    \maketitle

    \section{Introduction}

    QCD (Quantum Chromodynamics), the theory of the strong interaction among quarks and gluons, exhibits two main features: asymptotic freedom and color confinement. The thermodynamics of QCD matter is expressed in terms of the equation of state, which has been a subject of intense interest. Results from RHIC (Relativistic Heavy Ion Collider) have confirmed the existence of a deconfined state of matter that can be described in terms of quark and gluon degrees of freedom and behaves like a nearly ideal, strongly interacting fluid known as the quark-gluon plasma or, alternatively, the partonic phase at high temperature $T$ \cite{Gyulassy:2004zy}. At low temperature and baryon chemical potential $\mu$, the QCD matter is in the hadronic state where quarks and gluons are confined in hadrons. The hadron resonance gas model is widely used to study this state of matter since the results from it have been found to be in good agreement with the first principle calculations from lattice QCD \cite{HotQCD:2014kol, Bazavov:2017dus}. It is of great importance to study the transition between the quark-gluon and hadronic phases at different values of $\mu$. The (2+1)-flavor lattice calculations with physical quark masses have shown that the transition at vanishing $\mu$ is a rapid crossover at $T = 155 - 160$ MeV \cite{Borsanyi:2020fev}. Many model calculations, such as the linear $\sigma$ model and Nambu--Jona-Lasinio models, predict the existence of a critical point where the crossover line turns into first-order phase transition line at large $\mu$ \cite{Klevansky:1992qe,Buballa:2003qv,Stephanov:2004wx,Scavenius:2000qd}.   This is in a regime inaccessible to lattice calculations due to the infamous sign problem. This suspected critical point has not yet been observed in experiments. One of the major goals of the Beam Energy Scan II (BES-II) at RHIC is to search for the critical point and pinpoint its location in the QCD phase diagram. The QCD equation of state with a critical point is a key ingredient for hydrodynamic simulations relevant to the experimental search for the critical point. The scaling and universality of the critical behavior allow us to make predictions about the QCD critical point and its signatures in heavy-ion collisions despite having only incomplete information about the QCD equation of state. Universality tells us that the properties of systems in the vicinity of a critical point depend solely on the dimensionality and symmetry of the systems. It is generally expected that the critical point of QCD belongs to the 3D Ising universality class \cite{Stephanov:1999zu}. The goal of this paper is to improve upon the constructions of the equation of state reported in 
    \cite{Kapusta:2022pny} which is consistent with (i) lattice QCD for all $T$ and small $\mu$, (ii) perturbative QCD
    for large $T$ and/or large $\mu$, and (iii) a critical point with critical exponents and amplitude ratios from the same universality class as the liquid-gas phase transition and the 3D Ising model.

    The outline of this paper is as follows. In \cref{sec:singular eos} we briefly summarize an approach \cite{Kapusta:2022pny} to embed a critical point into the QCD equation of state. This is based on work by Schofield \cite{schofield1969correlation}. In \cref{sec: including the background} we describe how to merge the critical part with the background equation of state. In \cref{sec: mux} we describe several options for determining the line of crossover and phase transition, including numerical results and comparisons to freeze-out data from heavy ion collisions. The conclusion is given in \cref{sec: conclusions}.

    
    \section{Embedding the critical point}
    \label{sec:singular eos}
    
    In this section we briefly review the approach of Ref. \cite{Kapusta:2022pny} to embed the critical point into a smooth background equation of state. This approach is based on the work of \Cite{widom1965equation, Schofield:1969zza, schofield1969correlation, Rehr:1973zz,Karsch:2023pga}, which parametrizes the critical, scaling part of the equation of state. Such a parametrization has drawn much attention and has been used by several groups to study the critical behavior of the QCD critical point \cite{Nonaka:2004pg, Parotto:2018pwx, Kahangirwe:2024cny}.
    
    Using Ising model notation in the critical region, the magnetization $M$, reduced temperature $t = (T-T_{c})/T_{c}$, and magnetic field $H$ are expressed in terms of the variables $R$ and $\theta$ according to
    \begin{equation}
        M = m_{0}R^{\beta}\theta, \quad t = R(1-\theta^{2}), \quad H = h_{0}R^{\beta\delta}
        h(\theta)
    \end{equation}
    where $m_{0}$ and $h_{0}$ are positive normalization constants. The variable $R$ is nonnegative and measures the distance from the critical point in the $(t,H)$ plane; the critical point is located at $R = 0$ where $t = H = M = 0$. The variable
    $\theta$ parametrizes the displacements along the lines of constant $R$. For the 3D Ising model
    \begin{equation}
        \label{eq: h theta}h(\theta) = \theta (1 + h_{3}\theta^{2}+ h_{5}\theta^{4}
        )
    \end{equation}
    with $h_{3}= -0.762$, $h_{5}= 0.008$, and the critical exponents are $\beta = 0.3264$, $\gamma = 1.2371$. If we denote the smallest positive zero of $h(\theta)$ as $\theta_{0}$, then the range of $\theta$ is $-\theta_{0}\leq \theta \leq \theta_{0}$. With the specific choice of $h_{3}$ and $h_{5}$, $\theta_{0}\approx 1.154$. The coexistence curve corresponds to $\theta = \pm \theta_{0}$, where $H = 0$ and $M_{\pm}= \pm m_{0}R^{\beta}\theta_{0}$. 
    
    The studies by Refs. \cite{Nonaka:2004pg, Parotto:2018pwx, Kahangirwe:2024cny} use two maps, one of which is not globally invertible, to convert from Ising model variables to QCD quantities such as pressure and chemical potential. This introduces unnecessary parameters and theoretical uncertainties.  Instead, one can use any pair of thermodynamic variables conjugate to each other, as pointed out in \Cite{KUMAR198357,Pelissetto:2000ek}.  Here we follow Ref. \cite{Kapusta:2022pny} and write
    \begin{equation}
        \frac{n-n_{\text{c}}}{n_{\text{c}}}= m_{0}R^{\beta}\theta, \quad \frac{\mu
        - \mu_{x}(T)}{\mu_{\text{c}}}= h_{0}R^{\beta\delta}h(\theta)
    \end{equation}
    to explicitly include the phase coexistence boundary described by the function $\mu_{x}(T)$. We shall discuss how to determine $\mu_{x}(T)$ in \cref{sec: mux}, which is the main focus of this paper.

    The pressure satisfies the condition $(\partial P/\partial \mu)_{T}= n$ and is given by
        \begin{equation}
        P - P_{c} = \left[ \mu(R,\theta) - \mu_{c} \right] n(R,\theta) - m_{0} h_{0}
        \mu_{c} n_{c} R^{2-\alpha}g(\theta)
    \end{equation}
    The function $g(\theta)$ involves no new parameters and is given in the aforementioned references.  The corresponding Helmholtz free energy is
    \begin{equation}
        \label{eq: helmholtz free energy}
        f = \mu_{c} n + m_{0} h_{0} \mu_{c} n_{c} R^{2-\alpha}g(\theta) - P_{c}
    \end{equation}
    Given the free energy, one can then find the critical behavior. 
    
    The next step is to merge the singular equation of state into a smooth background equation of state in the $T$ and $\mu$ plane. It can be achieved by writing the pressure as
    \begin{equation}
        \label{eq: Schofield eos}P(\mu,T) = P_{\text{BG}}(\mu,T) + W(\mu,T) P_{*}(R,\theta)
    \end{equation}
    where $P_{\text{BG}}(\mu,T)$ is a smooth function of $T$ and $\mu$ and
    \begin{equation}
        P_{*}(R,\theta) = P_{0} + h_{0} \mu_{c} n_{0} R^{\beta\delta}h + m_{0} h_{0}
        \mu_{c} n_{0} R^{2-\alpha}\left[ \theta h(\theta) - g(\theta)\right]
    \end{equation}
    is the contribution from the critical part of the equation of state. The window function is chosen to be
    \begin{equation}
        W(\mu,T) = \exp\left[ -\left( \frac{\mu^{2j}- \mu_{x}^{2j}(T)}{c_{*} \mu_{c}^{j}
        \mu^{j}}\right)^{2} \right]
    \end{equation}
    where $j$ is a positive integer and $c_{*}$ is a number that controls the extent of the critical region that is included to make the singular part even in $\mu$, and vanishes in the vacuum $T = n = 0$. It introduces no extra singularity.  However, there exists a small residual effect of the critical point above $T_{c}$, and such a residual effect, if desired, can be reduced by modifying the window function in such a way that it decreases with temperature when $T > T_{c}$. One option, as given in \Cite{Kapusta:2022pny}, is to add an extra factor of 
    \begin{equation}
    1 - \exp{\left[ -(t_0/t)^2\right]}
    \label{eq: window function factor}
    \end{equation}
    to the window function for $t > 0$ with $t_{0} = 0.15$.

    The bulk thermodynamic quantities can be derived from the equation of state as constructed. For details see Ref. \cite{Kapusta:2022pny}.  For $T > T_{c}$, the equation of state and its derivatives are smooth, hence no true phase transition occurs. In contrast, for $T < T_{c}$, the thermodynamic quantities $P_{*}$, $s_{*}$, and $n_{*}$ exhibit discontinuities at $\mu = \mu_{x}$ due to the non-analyticity of $\theta$. These discontinuities determine a phase boundary where a first-order phase transition takes place.

    There are eight free parameters: $T_{c}$, $\mu_{c}$, $P_{0}$, $h_{0}$, $m_{0}$, $n_{0}$, $c_{*}$ and $j$ that need to be set before we perform the numerical calculations. Of course, none of them are universal; any choice is feasible as long as it leads to reasonable thermodynamics. We take $T_{c}$ and $\mu_{c}$ as the most interesting and fundamental. From now on, for heuristic reasons, we choose $P_{0} = 0.05 P_{c}$, $n_{0} = 0.1 n_{c}$, $m_{0} = 0.5$, $h_{0} = 0.2$, $j=2$, and $c_{*} = 1.0$.
    
    
    \section{Including the background}
    \label{sec: including the background}
    
    The background pressure $P_{\text{BG}}$ in \cref{eq: Schofield eos} needs to be specified. Any physically reasonable equation of state may be used so long as it fits the lattice results.  The approach taken in this paper does not depend on any particular one. For concreteness, we use the pressure of excluded volume model I (exI) and the pressure obtained from perturbative QCD \cite{Albright:2014gva,*Albright:2015uua}.  A switching function $S(\mu,T)$, which ranges between 0 and 1, is used to determine the pressure contribution of each phase. The background pressure $P_{\text{BG}}$ in 
    \cref{eq: Schofield eos} is taken to be
    \begin{equation}
    \label{eq: background eos}
        P_{\text{BG}}(T,\mu)= S(T,\mu) P_{q}(T,\mu) + \left[ 1 - S(T,\mu) \right]
        P_{h}(T,\mu)
    \end{equation}
    The switching function is modified from \cite{Albright:2014gva,*Albright:2015uua} to be
    \begin{equation}
     S(T, \mu) = \exp \left(- M_0^4/M^4\right)
    \end{equation}
    where
    \begin{equation}
    M = \sqrt{(\pi C_T T)^2 + (C_\mu \mu_q)^2}
    \label{M}
    \end{equation}
    is the renormalization scale for $\alpha_s$ in perturbative QCD.  This naturally connects the scale of the switching function with the strength of the QCD running coupling.
We use the 3-loop running coupling coupling from the PDG \cite{Workman:2022ynf} with one modification to $t$ (not the same $t$ as defined above but there ought to be no confusion from the context): we introduce a constant $C_S$ to soften its divergence at low temperatures and chemical potentials.
\be
 t \equiv \ln \left(C_S^2 + M^2/\Lambda^2_{\overline{MS}} \right)
\ee
With no loss of generality we take $\Lambda_{\overline{MS}} = 290$ MeV.

\begin{figure}[H]
    \centering
    \begin{minipage}[b]{0.49\textwidth}
        \includegraphics[width=\textwidth]{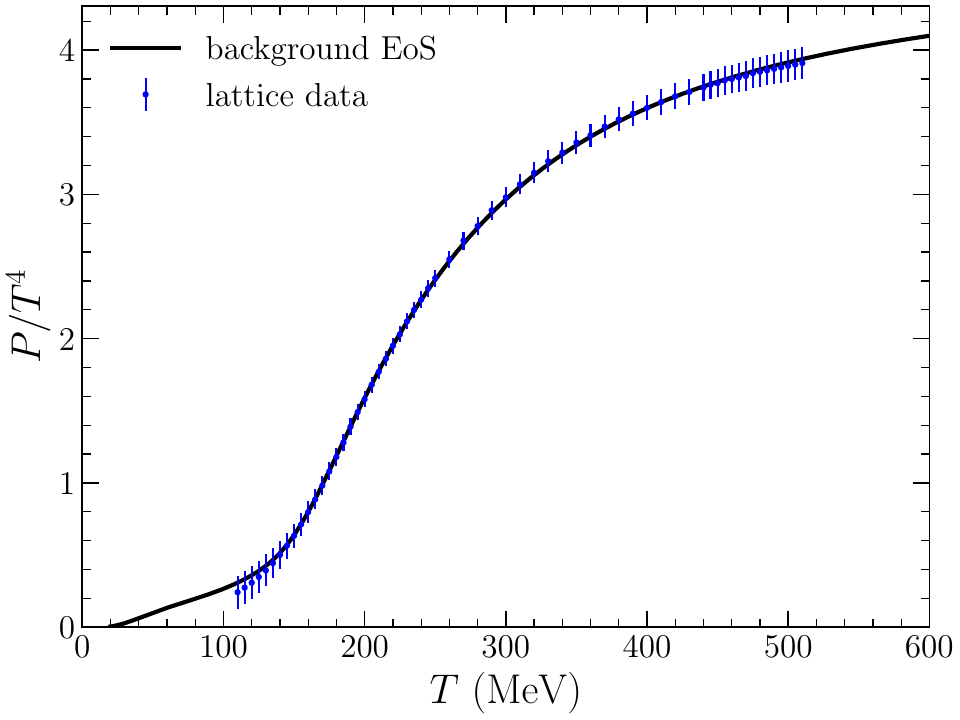}
        \caption{Pressure versus temperature.}
        \label{fig:fitting lattice pressure}
    \end{minipage}
    \hfill
    \begin{minipage}[b]{0.49\textwidth}
        \includegraphics[width=\textwidth]{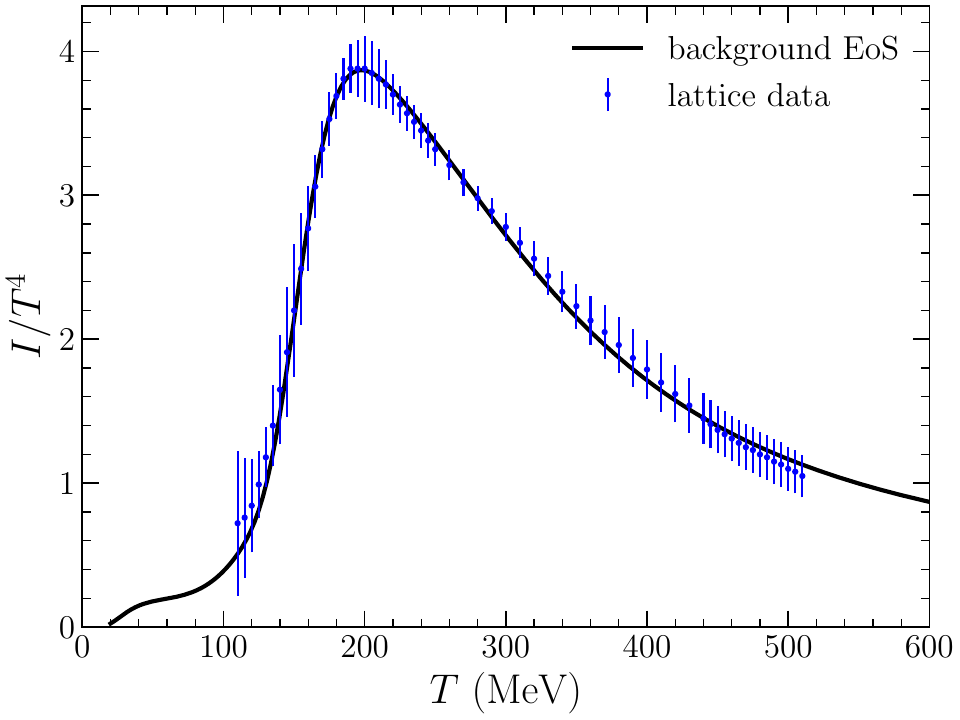}
        \caption{Trace anomaly versus temperature.}
        \label{fig:fitting lattice trace anomaly}
    \end{minipage}
\end{figure}

Using \cref{eq: background eos}, we can calculate the pressure and trace anomaly, also known as the interaction measure, at vanishing baryon chemical potential, that is $P_{\text{BG}}(T,\mu=0)$ and $I_{\text{BG}}(T,\mu=0) = \epsilon_{\text{BG}}(T,0) - 3P_{\text{BG}}(T,\mu=0)$, respectively.  They are then used to fit the lattice results. The fits are shown in \cref{fig:fitting lattice pressure} and \cref{fig:fitting lattice trace anomaly}.  This leads to the values $C_S=4.55$, $C_T = 3.67$, $M_0 = 2283.22$ MeV, and $\epsilon_0 = 1732.18$ MeV.  The combined $\chi^2$ per degree of freedom is 0.188.

\begin{figure}[H]
    \centering
    \begin{minipage}[b]{0.49\textwidth}
        \includegraphics[width=\textwidth]{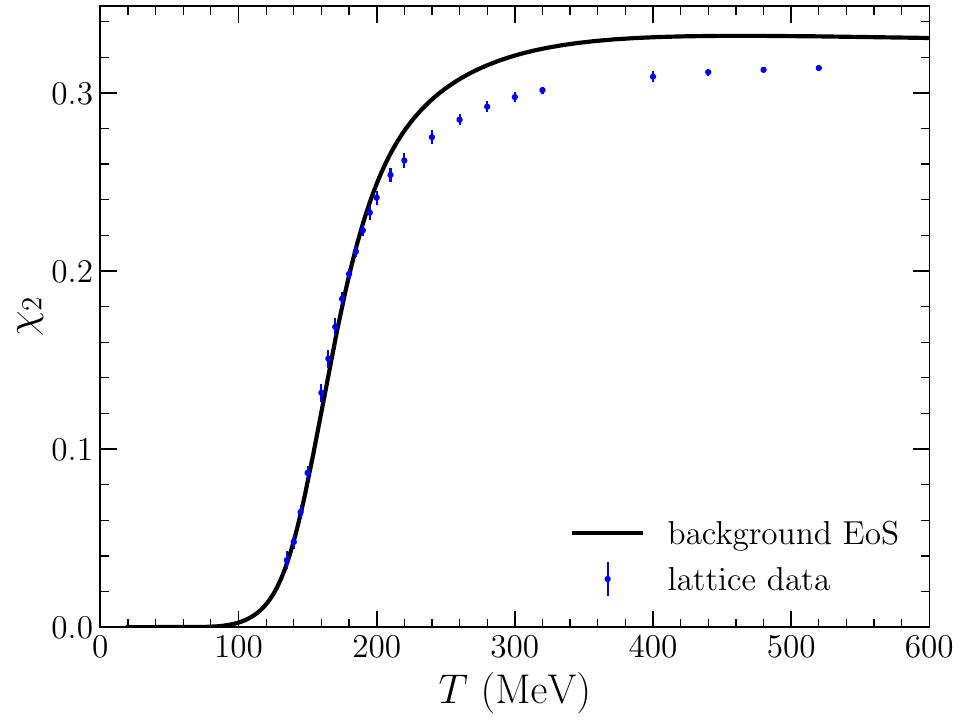}
        \caption{$\chi_2$ versus temperature.}
        \label{fig:fitting lattice chi2}
    \end{minipage}
    \hfill
    \begin{minipage}[b]{0.49\textwidth}
        \includegraphics[width=\textwidth]{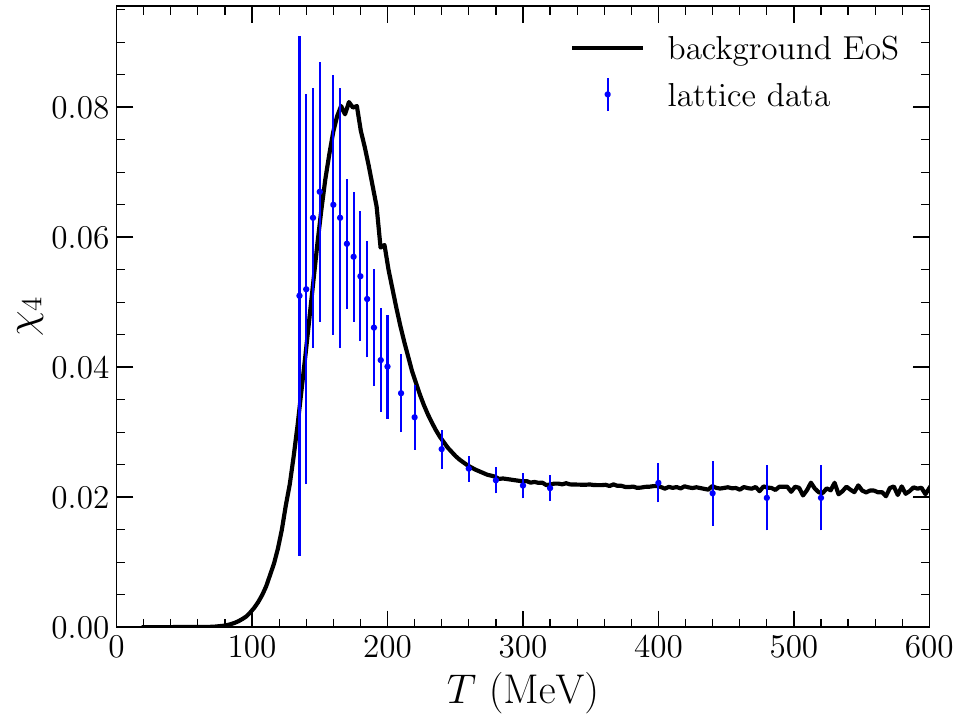}
        \caption{$\chi_4$ versus temperature.}
        \label{fig:fitting lattice chi4}
    \end{minipage}
\end{figure}

We then utilize the lattice results for $\chi_2 \equiv \partial^2(P/T^4)/\partial(\mu/T)^2$ at $T=0$ to determine the optimal value of $C_\mu$.  This is shown in \cref{fig:fitting lattice chi2} with $C_\mu = 0.9$. The $\chi_2$ so calculated deviates from the lattice results when $T>200$ MeV due to the perturbative QCD contributions. \cref{fig:fitting lattice chi4} shows the comparison for $\chi_4 \equiv \partial^4(P/T^4)/\partial(\mu/T)^4$ using the previously determined parameters.

This background equation of state does not include realistic nuclear forces which are important for cold or cool nuclear matter.  In particular, it does not include the nuclear liquid-gas phase transition which has a critical temperature of about 16 MeV and a critical density of about 0.05 fm$^{-3}$.  Therefore we have only shown results for $T \ge 20$ MeV.

    \section{Determining the phase transition curve and crossover line}
    \label{sec: mux}

    Whether or not there is a line of first-order phase transition ending in a critical point at $T_c$ and $\mu_c$ in the QCD phase diagram is unknown, let alone the shape of this line described by the function $\mu_x(T)$.  That function is a crucial ingredient to the approach described above as well as others.  Reference \cite{Kapusta:2022pny} determined it via the condition
    $n_{BG}(\mu_x(T),T) = n_c - n_0$ when $T \le T_c$. The reason for this choice is that it results in a symmetric, inverted U-shaped curve in the $T-n$ plane as observed in some atomic and molecular liquid-gas phase transitions. In this paper we take $T_{c} = 120$ MeV and $\mu_{c} = 670$ MeV.  This results in the critical density $n_c = 0.2957$ fm$^{-3}$.  Reasons for this particular choice are given below.  
    \begin{figure}[h!]
        \centering
        \begin{minipage}[b]{0.49\textwidth}
            \includegraphics[width=\textwidth]{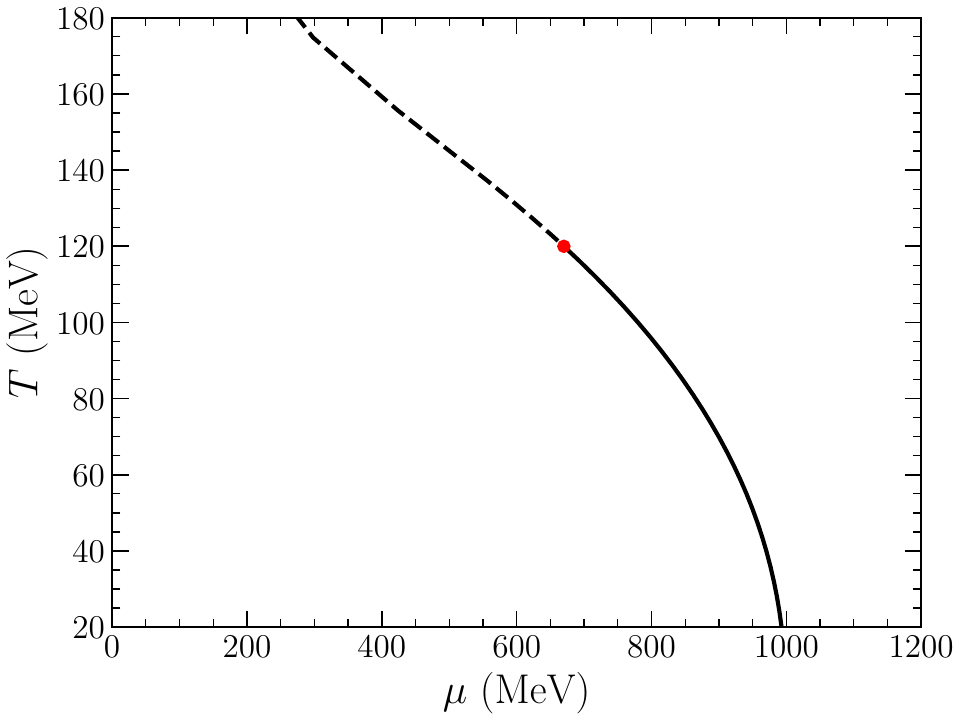}
            \caption{The phase coexistence boundary determined by the background density following Ref. \cite{Kapusta:2022pny}.  The dashed curve is the analytic extension for $T > T_c$.}
            \label{fig:original mu}
        \end{minipage}
        \hfill
        \begin{minipage}[b]{0.49\textwidth}
            \includegraphics[width=\textwidth]{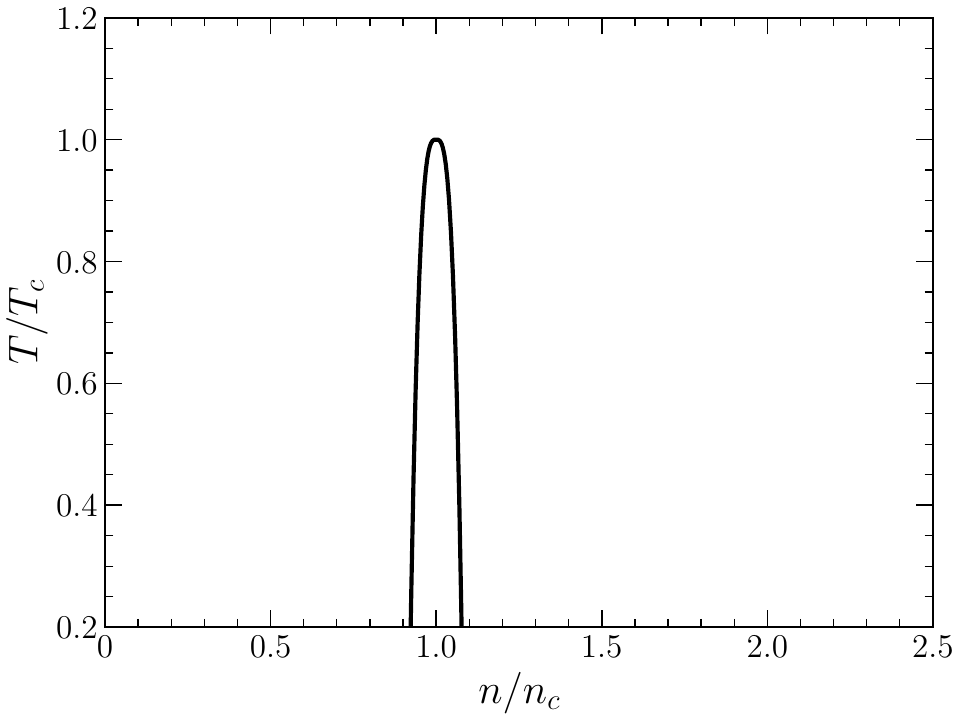}
                        \caption{The phase coexistence boundary determined by the background density following Ref. \cite{Kapusta:2022pny}.}
                        \vspace{10pt}
            \label{fig:original n}
        \end{minipage}
    \end{figure} 
    Figure \ref{fig:original mu} shows $\mu_x(T)$ versus $T$.   Figure \ref{fig:original n} shows the phase coexistence in the temperature versus density plane.  As a consequence of the isochor condition, the analytic extension of $\mu_x(T)$ for $T > T_c$ never intersects the temperature axis.  Intuitively, it might be expected that this analytic extension would intersect the temperature axis where lattice calculations show a peak in the heat capacity signaling a near phase transition.
    
    Another widely applied scenario is to map the phase diagram of the 3D Ising model onto QCD through a linear transformation of the thermodynamic variables \cite{Parotto:2018pwx,Pradeep:2024cca,Karthein:2024zvs}.  In that approach a Taylor expansion in powers of $\mu/T$ is performed about $\mu = 0$.  Terms up to a finite order are reshuffled between the background lattice equation of state, calculated at $\mu = 0$, and the critical equation of state.  Then a symmetrization is done to ensure that the pressure is an even function of $\mu$.  That procedure limits how large $\mu/T$ can be before unphysical behavior is manifest in the equation of state.  Determination of $\mu_x(T)$ is intimately entwined with that construction. It has also been argued quite successfully that the thermal conditions at the time of chemical freeze-out in heavy ion collision can be characterized by lines in the $T-\mu$ plane on which certain thermodynamic observables or ratios thereof stay constant \cite{Cleymans:1999st,Cleymans:2005xv}. Moreover, lattice calculations show that when the baryon chemical potential $\mu$ is relatively small, these approximately agree with the crossover line in (2+1)-flavor QCD \cite{Bazavov:2017dus, HotQCD:2018pds}.

    In this paper we follow the philosophy of Ref. \cite{Kapusta:2022pny} but modify the condition which determines $\mu_x(T)$ so that the crossover line from the critical point intersects the temperature axis.  For example, consider determining the crossover curve based on the entropy density instead of the baryon density.  This curve would intersect the temperature axis but not the chemical potential axis.  A natural possibility then would be to consider the sum of the squares, namely
    \begin{equation}
        \label{eq: constant square linear combination condition}
        \left( \frac{\tilde{s}}{\tilde{s}_{c}}
        \right)^{2} + \left( \frac{\tilde{n}}{\tilde{n}_{c}}\right)^{2} = 2 \quad \text{Condition A}
    \end{equation}
    where $\tilde{s} = s_{\text{BG}}(T,\mu_{x}(T))$, $\tilde{s}_c = s_{\text{BG}}(T_c,\mu_{x}(T_c))$, and similarly for the density $\tilde{n}$. The function $\mu_x(T)$ will then intersect both the temperature and chemical potential axes, and the resulting equation of state will be an even function of $\mu$, as it should be.  We call this condition A and, when an ambiguity may occur, we label thermodynamic variables or functions with the superscript $A$.  A second natural possibility to determine 
    $\mu_x(T)$ is via the energy density
    \begin{equation}
        \label{eq: constant energy density condition}
        \frac{\tilde{\epsilon}}{\tilde{\epsilon}_{c}} = 1 \quad \text{Condition B}
    \end{equation}
    We call this condition B and label thermodynamic variables or functions with the superscript $B$.  Similar to condition A, the function $\mu_x(T)$ will intersect both the temperature and chemical potential axes and the resulting equation of state will be an even function of $\mu$.

    \begin{figure}[h!]
        \centering
        \begin{minipage}[b]{0.49\textwidth}
            \includegraphics[width=\textwidth]{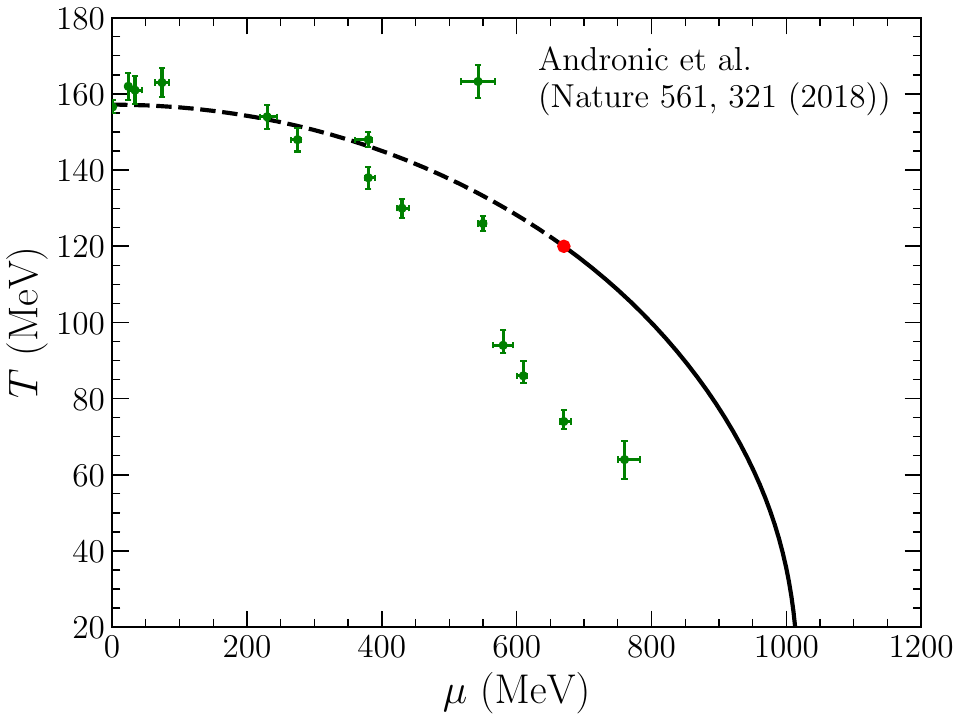}
            \caption{The crossover (dashed) and first-order (solid) transition lines determined by condition A. The data points represent chemical freeze-out points that result from a statistical hadronization analysis of hadron yields for central collisions at different energies \cite{Andronic:2017pug}.}
            \label{fig:data A}
        \end{minipage}
        \hfill
        \begin{minipage}[b]{0.49\textwidth}
            \includegraphics[width=\textwidth]{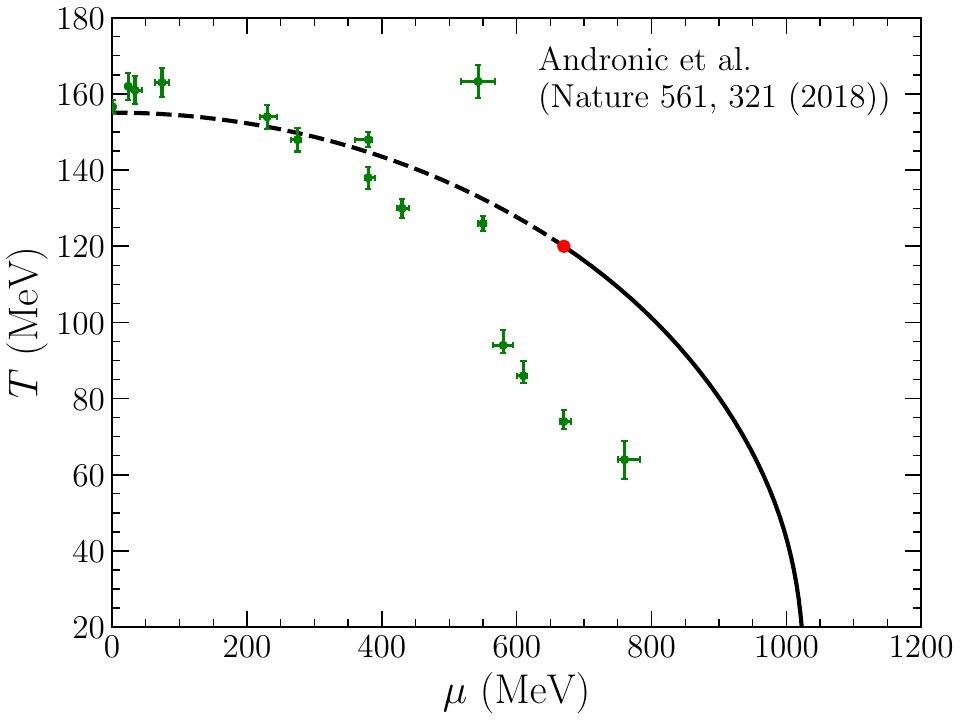}
            \caption{Same as Fig. \ref{fig:data A} for condition B.
            \vspace{45pt}}
            \label{fig:data B}
        \end{minipage}
    \end{figure}
    The resulting $\mu_{x}^{A}(T)$ and $\mu_{x}^{B}(T)$ are shown in Figs. \ref{fig:data A} and \ref{fig:data B}. Also shown are the chemical freeze-out points obtained from a single, consistent statistical hadronization analysis of hadron yields for central collisions at different energies \cite{Andronic:2017pug}.  References to the original experimental data are cited in that paper.  A critical point at $T_{c} = 120$ MeV and $\mu_{c} = 670$ MeV was chosen so that the crossover curve, the dashed line, passed through the data points; a $\chi^2$ fit was not done, only a fit by eye. This is suggestive, but not proof, of the location of a critical point. The crossover curves approximately agree with the chemical freeze-out points at low $\mu$ while the critical curves at large $\mu$ are noticeably above these points; both qualitative behaviors are consistent with previous studies \cite{Pasztor:2024dpv,Braun-Munzinger:2003htr,Flor:2020fdw, Floerchinger:2012xd}.  It might also be due to a change in the dynamics in heavy ion collisions at the beam energy decreases from high to medium to low.

\begin{figure}[h!]
        \centering
        \begin{minipage}[b]{0.49\textwidth}
            \includegraphics[width=\textwidth]{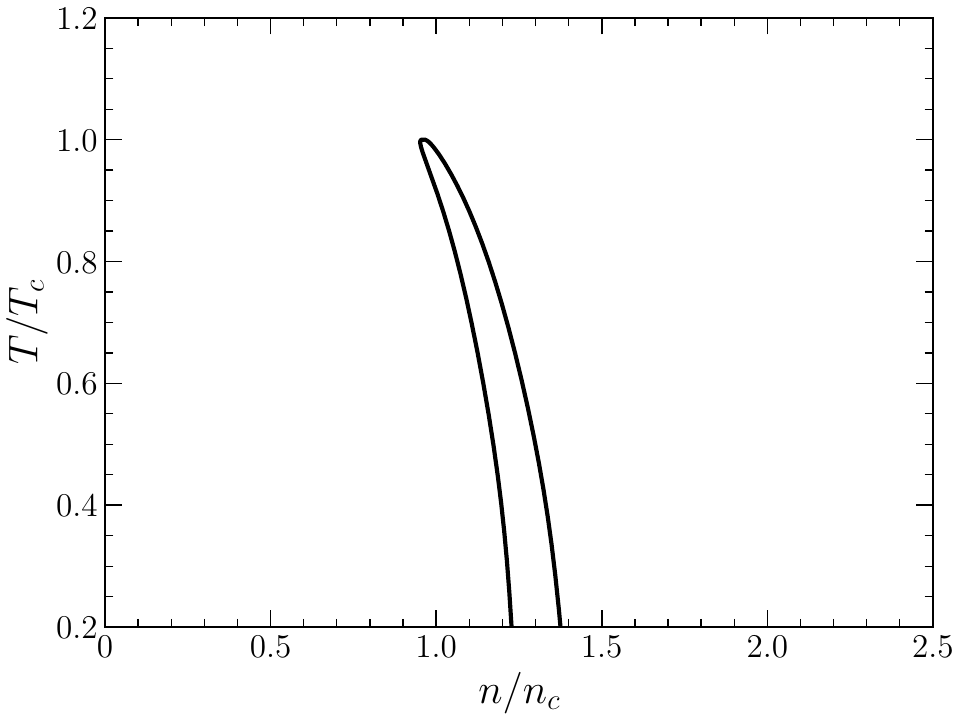}
            \caption{Coexistence curve in the temperature versus density plane for $\mu_x^A$.}
            \label{fig:coexist A}
        \end{minipage}
        \hfill
        \begin{minipage}[b]{0.49\textwidth}
            \includegraphics[width=\textwidth]{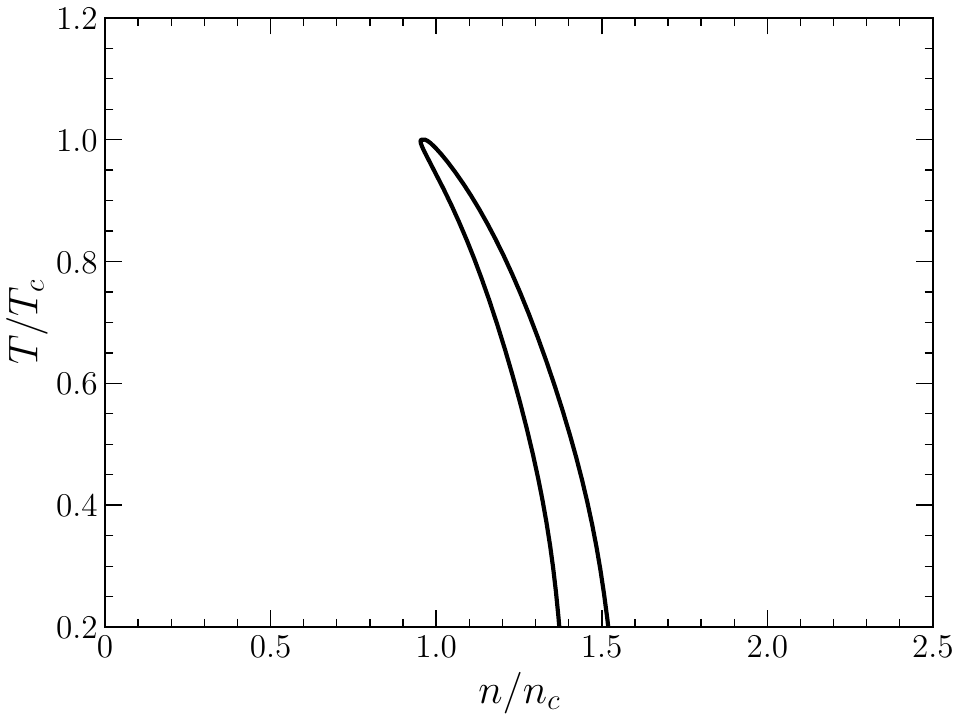}
            \caption{Coexistence curve in the temperature versus density plane for $\mu_x^B$.}
            \label{fig:coexist B}
        \end{minipage}
    \end{figure}
    The coexistence curves in the $T - n$ plane resulting from conditions A and B are shown in Figs. \ref{fig:coexist A} and \ref{fig:coexist B}. The curves no longer have an inverted U shape.  The fact that $n_{\text{BG}}$ along the critical curve now increases as $\mu$ increases results in right-skewed coexistence curves.
    
        \begin{figure}[h!]
        \centering
        \begin{minipage}[b]{0.49\textwidth}
            \includegraphics[width=\textwidth]{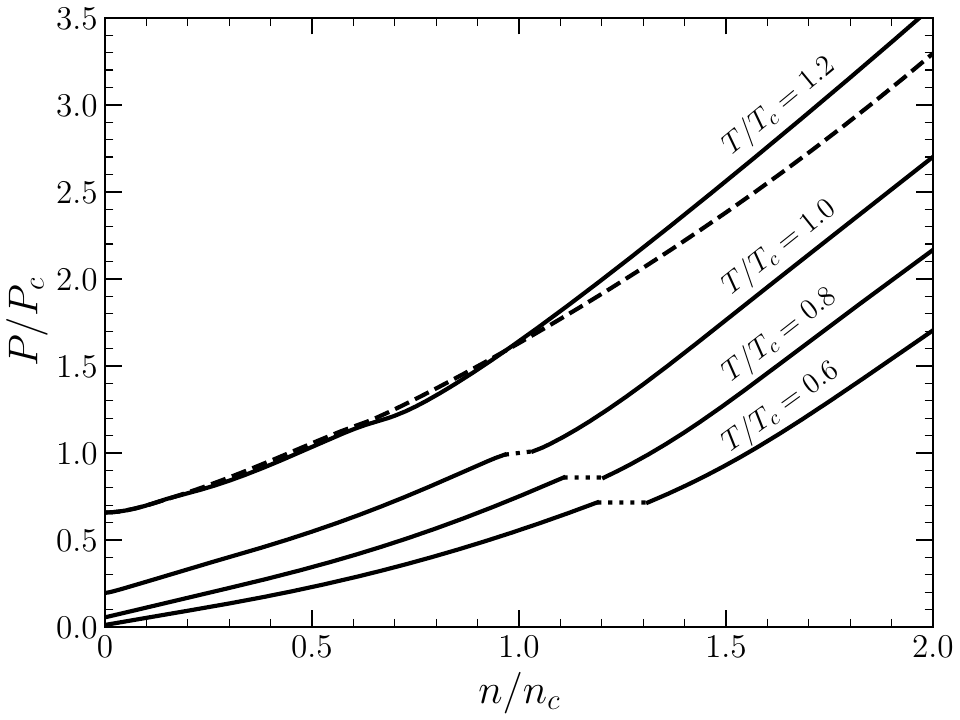}
            \caption{Isotherms of pressure versus density for $\mu_x^A$.  The dashed curve for $T/T_c = 1.2$ includes the factor of (\ref{eq: window function factor}) while the solid curve does not.}
            \label{fig:isotherm A}
        \end{minipage}
        \hfill
        \begin{minipage}[b]{0.49\textwidth}
            \includegraphics[width=\textwidth]{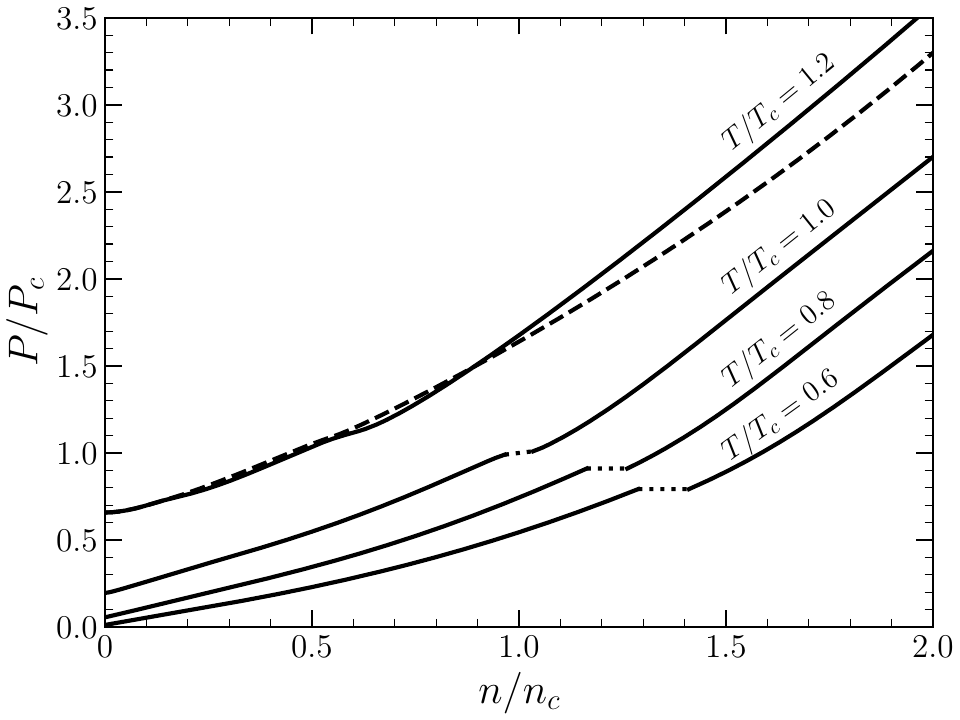}
            \caption{Isotherms of pressure versus density for $\mu_x^B$. The dashed curve for $T/T_c = 1.2$ includes the factor of (\ref{eq: window function factor}) while the solid curve does not.}
            \label{fig:isotherm B}
        \end{minipage}
    \end{figure}
    Figures \ref{fig:isotherm A} and \ref{fig:isotherm B} show the isotherms of pressure versus density for $T/T_c= 0.6, 0.8, 1.0$, and $1.2$. The residual effect of the critical point above $T_c$ is eliminated by including the extra factor given in \cref{eq:
    window function factor} with the window function; the dashed lines in the upper panels are the corresponding isotherm.
    
\begin{figure}[!ht]
        \centering
        \begin{minipage}[b]{0.49\textwidth}
            \includegraphics[width=\textwidth]{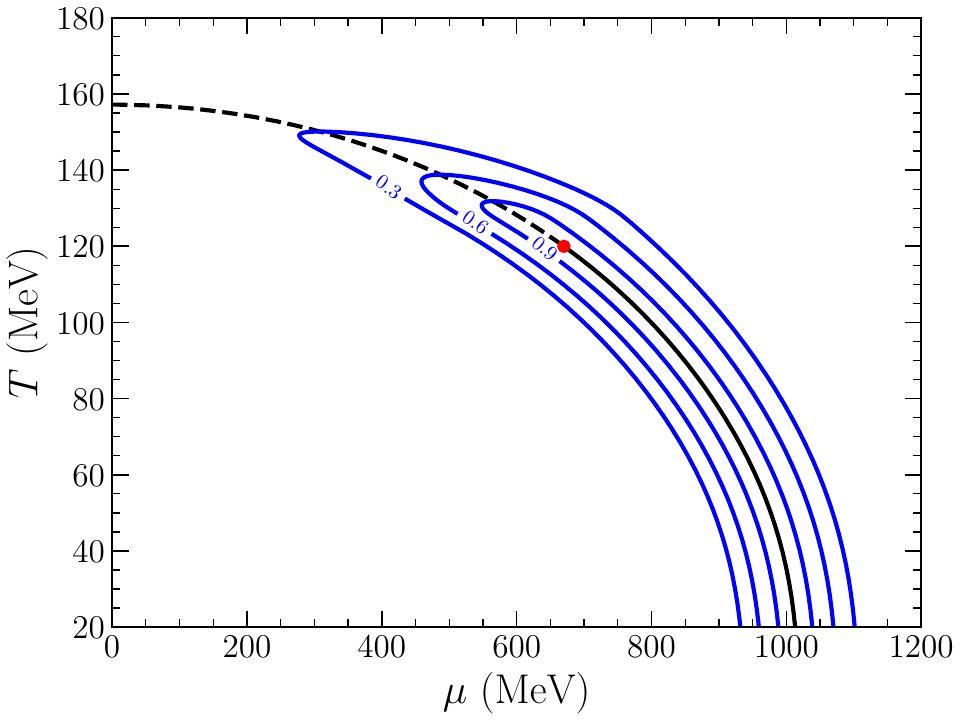}
            \caption{Window function contours for $\mu_x^A$ with the factor of (\ref{eq: window function
        factor}).}
            \label{fig:contour A}
        \end{minipage}
        \hfill
        \begin{minipage}[b]{0.49\textwidth}
            \includegraphics[width=\textwidth]{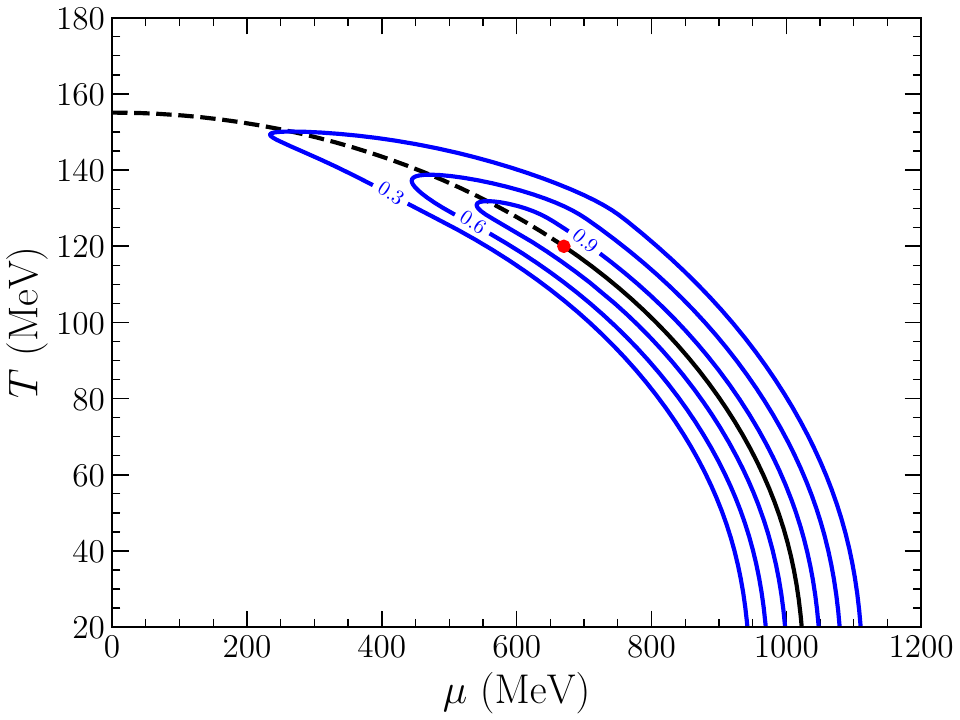}
            \caption{Window function contours for $\mu_x^B$ with the factor of (\ref{eq: window function
        factor}).}
            \label{fig:contour B}
        \end{minipage}
    \end{figure}

    The contours of the window function with the extra factor in the $T$ versus $\mu$ plane at levels 0.3, 0.6, and 0.9 are depicted in Figs. \ref{fig:contour A} and \ref{fig:contour B}.  The parameters in the window function determine the broadness of the phase transition region.

    \begin{figure}[!ht]
        \centering
        \begin{minipage}[b]{0.49\textwidth}
            \includegraphics[width=\textwidth]{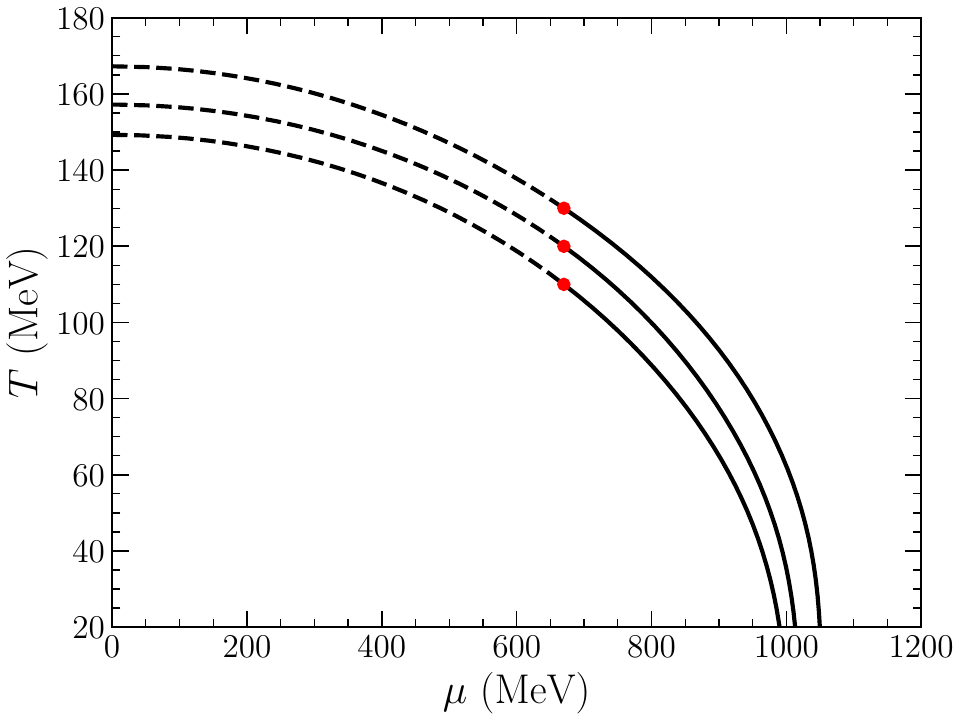}
            \caption{How the phase boundary and crossover curves change with $T_c$ using condition A.}
            \label{fig:vary T_c}
        \end{minipage}
        \hfill
        \begin{minipage}[b]{0.49\textwidth}
            \includegraphics[width=\textwidth]{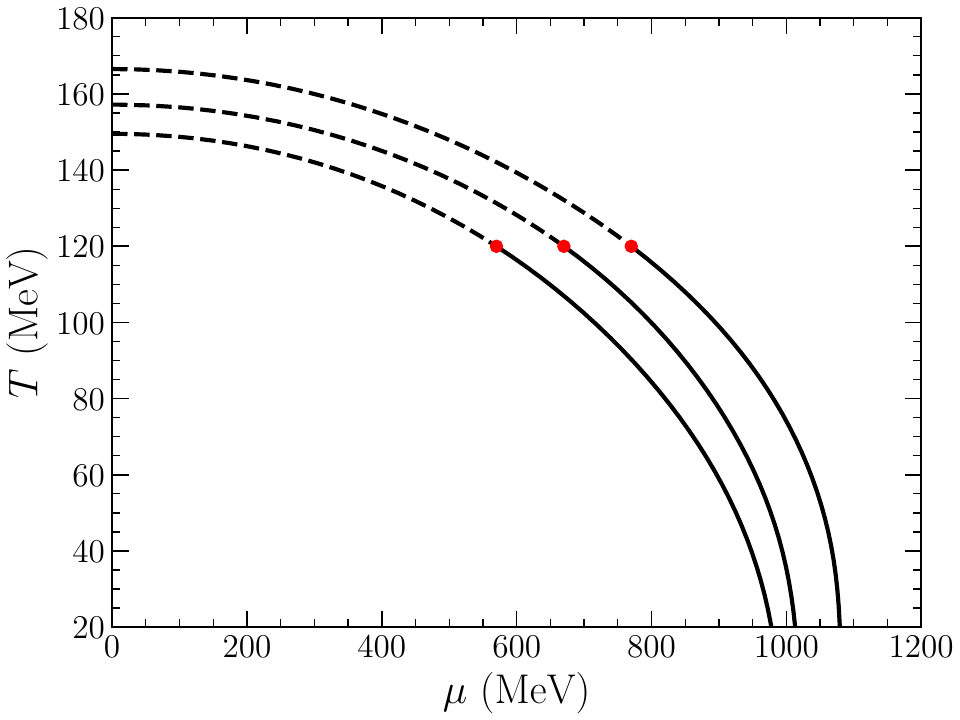}
            \caption{How the phase boundary and crossover curves change with $\mu_c$ using condition A.}
            \label{fig:vary mu_c}
        \end{minipage}
    \end{figure}
    Finally, we investigate the sensitivity of the phase boundary and crossover curve by changing $T_c$ and 
    $\mu_c$ while keeping all other parameters fixed. Some results are shown in Figs. \ref{fig:vary T_c} and \ref{fig:vary mu_c}.  The curves move up and down in the expected manner.


    \newpage
   
    \section{Conclusions}
    \label{sec: conclusions}
    In this paper we briefly reviewed the approach introduced in \Cite{Kapusta:2022pny} to embed a critical point into a smooth background equation of state. This is based on the widely used assumption that the QCD critical point, if it exists, belongs to the 3D Ising and liquid-gas universality class. The condition employed in \Cite{Kapusta:2022pny} leads to an inverted U-shaped phase coexistence curve in the temperature versus density plane. The analytic extension of the phase coexistence chemical potential $\mu_x(T)$ into the crossover region does not intersect with the $T$ axis as one might intuitively expect. As alternatives, we used two different conditions that lead to functions $\mu_x(T)$ that do agree with this expectation. The two 
    $\mu_x(T)$ so obtained are concave functions with globally negative derivatives. The location of a critical point is estimated  by fitting the crossover line with the chemical freeze-out points in central collisions of high energy heavy ions; it is suggestive rather than conclusive. The coexistence curves and the isotherms associated with the new phase boundaries were shown. We also investigated the sensitivity of the crossover curve and the first-order phase transition line by varying $T_c$ and $\mu_c$ separately. We find that the overall shape of the curves is insensitive to the precise location of the critical point.

    \newpage
    
    The goal of constructing an equation of state with a critical point is to implement it for hydrodynamic simulations of heavy ion collisions to investigate the existence of critical behavior. The equation of state can also be used in numerical simulations of neutron star mergers.  The construction is very flexible.  The location of the critical point, the critical exponents and amplitudes, the extent of the critical region, and the choice of the background equation of state can all be varied.
    
    \section{Acknowledgments}
    
We thank O. Chabowski, C. Plumberg and M. Singh for constructive feedback on the manuscript. This work was supported by the U.S. DOE Grant No. DE-FG02-87ER40328.

    \bibliography{reference}
\end{document}

%% file: abstract.tex
\begin{abstract}
    Lattice QCD calculations have shown that the transition from hadrons to quarks and gluons is a rapid crossover at $T = 155-160$ MeV at vanishing chemical potential. Many model calculations show that the transition is first-order at sufficiently high baryon chemical potential. It is then natural to expect the existence of a critical point where the crossover and first-order phase transition lines meet. We show how to embed a phase boundary that terminates at the critical point in a smooth background equation of state, using several different but closely related criteria, so as to yield the critical exponents and critical amplitude ratios expected of a transition in the 3D Ising and liquid-gas universality class. The crossover curves can be tuned to pass through experimental freeze-out data from heavy ion collisions at RHIC and the LHC.  The resulting equations of state can be used in hydrodynamic simulations of these collisions to probe the existence of a critical point and corresponding first-order phase transition.
\end{abstract}